%% file: arXiv_preprint.tex
\documentclass[preprint]{acm_proc_article-sp}

\usepackage{amsmath}
\usepackage{amssymb}
\usepackage[caption=false,font=footnotesize]{subfig}
\usepackage{url}
\usepackage{fancyhdr}
\usepackage{url}
\usepackage{listings}
\begin{document}

\title{Managing Communication Latency-Hiding at Runtime for Parallel Programming Languages and Libraries}
% on Distributed Memory Architectures
% NOTE: You are NOT restricted in how many productivity'rows' of

\numberofauthors{2} %  in this sample file, there are a *total*
\author{
% 1st. author
\alignauthor
Mads Ruben Burgdorff Kristensen\\
       \affaddr{Niels Bohr Institute}\\
       \affaddr{Copenhagen, Denmark}\\
       \email{madsbk@nbi.dk}
% 2nd. author
\alignauthor
Brian Vinter\\
       \affaddr{Niels Bohr Institute}\\
       \affaddr{Copenhagen, Denmark}\\
       \email{brian.vinter@nbi.dk}
}

% \toappear{PREPRINT To appear in the Ninth AES Conference on Medievil Lithuania  Embalming Technique, June 1991, Alfaretta, Georgia.}

\maketitle

\thispagestyle{fancy}
\rhead{\it PREPRINT}

\begin{abstract}
This work introduces a runtime model for managing communication with support for latency-hiding. The model enables non-computer science researchers to exploit communication latency-hiding techniques seamlessly. For compiled languages, it is often possible to create efficient schedules for communication, but this is not the case for interpreted languages. By maintaining data dependencies between scheduled operations, it is possible to aggressively initiate communication and lazily evaluate tasks to allow maximal time for the communication to finish before entering a wait state. We implement a heuristic of this model in DistNumPy, an auto-parallelizing version of numerical Python that allows sequential NumPy programs to run on distributed memory architectures. Furthermore, we present performance comparisons for eight benchmarks with and without automatic latency-hiding. The results shows that our model reduces the time spent on waiting for communication as much as 27 times, from a maximum of 54\% to only 2\% of the total execution time, in a stencil application.
\end{abstract}

\section{Introduction}
\input{intro}

\section{Related Work}
\input{relatedwork}

\input{theory}

\section{Distributed Numerical Python}
\input{distnumpy}

\subsection{Latency-Hiding}\label{sec:Latency-Hiding}
\input{latency_hiding}

\section{Experiments}
\input{benchmark}

\section{Future Work}\label{sec:futurework}
\input{futurework}

\section{Conclusions}
\input{conclusions}

\bibliographystyle{abbrv}
\bibliography{refs}

% ACM needs 'a single self-contained file'!
%
\balancecolumns
% That's all folks!
\end{document}

%% file: intro.tex
There are many ways to categorize scientific applications -- terms including scalability, communication pattern, IO and so forth. In the following, we wish to differentiate between large maintained codes, often commercial or belonging to a community, and smaller, less organized, codes that are used by individual researchers or in a small research group. The large codes are often fairly static and each version of the code can be expected to be run many times by many users, and thus justifying a large investment in writing the code. The small development codes on the other hand, change frequently and may only be run a few times after each change, usually only by the one user who made the changes. 

The consequence of these two patterns is that the large codes may be written in a compiled language with explicit message-passing. While the small codes have an inherent need to be written in a high-productivity programming language, where the development time is drastically reduced compared to a compiled language with explicit message-passing.

High-productivity languages such as Matlab and Python -- popular languages for scientific computing -- are generally accepted as being slower than compiles languages, but more importantly they are inherently sequential and while introducing parallelism is possible in these languages \cite{Hollingsworth96}\cite{PyCSP07}\cite{pupympi} it limits the productivity. It has been previously shown that it is possible to parallelize matrix and vector-based data-structures from Python, even on distributed memory architectures\cite{kristensen10_dnumpy}. However, the parallel execution speed is severely impeded by communication between nodes, in such a scheme for automatic parallelization.

%\subsection{Motivation}
To obtain performance in manual parallelization the programmer usually applies a technique known as latency-hid\-ing, which is a well-known technique to improve the performance and scalability of communication bound problems and is mandatory in many scientific computations. 

%We define latency-hiding informally as in \cite{Strumpen94latencyhiding} -- ``a technique to increase processor utilization by transferring data via the network while continuing with the computation at the same time''. When implementing latency-hiding the overall performance depends on two issues: the ability of the communication layer to handle the communication asynchronously and the amount of computation that can overlap the communication -- in this work we will focus on the latter issue.

In this paper, we introduce an abstract model to handle latency-hiding at runtime. The target is scientific applications that make use of vectorized computation. The model enables us to implement latency-hiding into high-produc\-tivity programming languages where the runtime system handles communication and parallelization exclusively.

In such high-productivity languages, a key feature is automatic distribution, parallelization and communication that are transparent to the user. Because of this transparency, the runtime system has to handle latency-hiding without any help. Furthermore, the runtime system has no knowledge of the communication pattern used by the user. A generic model for latency-hiding is therefore desirable.

The transparent latency-hiding enables a researcher that uses small self-maintained programs, to use a high-produc\-tivity programming language, Python in our case, without sacrificing the possibility of utilizing scalable distributed memory platforms. The purpose of the work is not that the performance of an application, which is written in a high-productivity language, should compete with that of a manually parallelized compiled application. Rather the purpose is to close the gap between high-productivity on a single CPU and high performance on a parallel platform and thus have a high-productivity environment for scalable architectures.

%\subsection{Target Programming Frameworks}
The latency-hiding model proposed in this paper is tailored to parallel programming languages and libraries with the following properties:
\begin{itemize}
\item The programming language requires dynamic scheduling at runtime because it is interpreted.
\item The programming language supports and utilizes a distributed memory environment.
\item All parallel processes have a global knowledge of the data distribution and computation.
\item The programming language makes use of data parallelism in a Single Instruction, Multiple Data (SIMD) fashion in the sense that data affinity dictates the distribution of the computation.
\end{itemize}
Distributed Numerical Python (DistNumPy) is an example of such a parallel library, and the first project that fully incorporate our latency-hiding model. The implementation of DistNumPy is open-source and freely available\footnote{DistNumPy is available at \url{http://code.google.com/p/DistNumPy}}.

The rest of the paper is organized as follows. In section 2, 3 and 4, we go through the background and theory of our latency-hiding model. In section 5, we describe how we use our latency-hiding model in DistNumPy. In section 6, we present a performance study. Section 7 is future work, and finally in section 8 we conclude.

%% file: relatedwork.tex
Libraries and programming languages that support parallelism in a high productive manner is a well-known concept. In a perfect framework, all parallelism introduced by the framework is completely transparent to the user while the performance and scalability achieved is optimal. However, most frameworks require the user to specify some kind of parallelism -- either explicitly by using parallel directives or implicitly by using parallel data structures. 

In this paper we will focus on data parallel frameworks, in which parallelism is based on the exploitation of data locality. A classical example of such a framework is High Performance Fortran (HPF) \cite{Loveman93}, which is an extension of the Fortran-90 programming language. HPF introduces parallelism primarily with vector operations, which, in order to archive good performance, must be aligned by the user to reduce communication. However, a lot of work has been put into eliminating this alignment issue either at compile-time or run-time \cite{Koelbel91} \cite{Brandes1994} \cite{Benkner97}.

%Partitioned Global Address Space (PGAS), a distributed shared memory model, is a group of languages that are designed around a memory model in which a global address space is partitioned such that a portion of it is local to each processor.
%Co-Array Fortran\cite{Co-arrayFortran98} is a small language extension of Fortran-95 for parallel processing on Distributed Memory Machines. It introduces PGAS by extending Fortran arrays with a \emph{co-array} dimension. Each process can access remote instances of an array by indexing into the co-array dimensions. A similar PGAS extension called Unified Parallel C (UPC)\cite{Carlson99_UPC} extents the C language with a distributed array declaration. Titanium\cite{Yelick1998_Titanium} is a PGAS dialect of Java that provides a global memory space abstraction where all data has a user-controllable processor affinity, but parallel processes may directly reference each others memory to read and write values. Common for all three languages is that the user explicitly expresses the parallelism and the data distribution in order to obtain good parallel performance. It is the responsibility of the compiler to find a good communication scheduling for the application.

DistNumPy\cite{kristensen10_dnumpy} is a library for doing numerical computation in Python that targets scalable distributed memory architectures. DistNumPy accomplishes this by extending the NumPy module\cite{numpy}, which combined with Python forms a popular framework for scientific computations. The parallelization introduced in DistNumPy focuses on parallel vector operations like HPF, but because of the latency-hiding we introduce in this paper, it is not a requirement to align vectors in order to achieve good performance.

Hardware architectures also exploit data parallelism to hiding memory latency \cite{Espasa97} or communication latency \cite{Kaxiras00}. Likewise, parallel data dependency analysis is essential in order to efficiently schedule instructions and avoid pipeline interlocks \cite{Hennessy83} \cite{Gibbons86}.

%% file: theory.tex
\section{Latency-Hiding}
We define latency-hiding informally as in \cite{Strumpen94latencyhiding} -- ``a technique to increase processor utilization by transferring data via the network while continuing with the computation at the same time''. When implementing latency-hiding the overall performance depends on two issues: the ability of the communication layer to handle the communication asynchronously and the amount of computation that can overlap the communication -- in this work we will focus on the latter issue.

In order to maximize the amount of communication hidden behind computation when performing vectorized computations our abstract latency-hiding model uses a greedy algorithm. The algorithm divides the arrays, and thereby the computation, into a number of fixed-sized data blocks. Since most numerical applications will work on identical dimensioned datasets, the distribution of the datasets will be identical. For many data blocks, the location will therefore be the same and these will be ready for execution without any data transfer. While the co-located data blocks are processed, the transfers of the data blocks from different location can be carried out in the background thus implementing latency-hiding.
The performance of this algorithm relies on two properties: 
\begin{itemize}
\item The number of data blocks must be significantly great\-er than number of parallel processors.
\item A significat number of data blocks must share location.
\end{itemize}
In order to obtain both properties we need a data structure that support easy retrieval of dependencies between data blocks. Furthermore, the number of data blocks in a computation is proportional with the total problem size thus efficiency is of utmost importance.

\section{Directed Acyclic Graph}\label{sec:theory}
It is well-known that a directed acyclic graph (DAG) can be used to express dependencies in parallel applications\cite{AhSeUl86}. Nodes in the DAG represent operations and edges represent serialization dependencies between the operations, which in our case is due to conflicting data block accesses.

Scheduling operations in a DAG is a well-studied problem. The scheduling problem is NP-complete in its general forms \cite{Garey1979} where operations are scheduled such that the overall computation time is minimized. There exist many heuristic for solving the scheduling problem \cite{Khan94}, but none match our target.

The scheduling problem we solve in this paper is not NP-hard because we are targeting programming frameworks that make use of data parallelism in a SIMD fashion. The parallel model we introduce is statically orchestrating data distribution and parallelization based on predefined data affinity. Assignment of computation tasks are not part of our scheduling problem. Instead, our scheduling problem consists of maximizing the amount of communication that overlaps computation when moving data to the process that is predefined to perform the computation.

In \cite{Song09} the authors demonstrate that it is possible to dynamic schedule operations in a distributed environment using local DAGs. That is, each process runs a private runtime system and communicates with other processes regarding data dependences. Similarly, our scheduling problem is also dynamic but in our case all processes have a global knowledge of the data distribution and computation. Hence, no communication regarding data dependences is required at all.

The time complexity of insetting a node into a DAG, $G=(V,E)$, is $O(V)$ in worse case. Building the complete DAG is therefore $O(V^2)$. Removing one node from the DAG is $O(V)$, which means that in the case where we simply wants to schedule all operations in a legal order the time complexity is $O(V^2)$. This is without minimizing the overall computation or the amount of communication hidden behind computation. We therefore conclude that a complete DAG approach is inadequate for runtime control of latency-hiding in our case.

We address the shortcoming of the DAG approach through a heuristic that manage dependencies on individual blocks. Instead of having a complete DAG, we maintain a list of depending operations for each data block. Still, the time complexity of scheduling all operations is $O(V^2)$ in worse case, but the heuristic exploits the observation that in the common case a scientific application spreads a vectorized operation evenly between the data blocks in the involved arrays. Thus the number of dependencies associated with a single data block is manageable by a simple linked list. In Section \ref{sec:dep_impl}, we will present a practical implementation of the idea.

%% file: distnumpy.tex
The programming language Python combined with the numerical library NumPy\cite{numpy} has become a popular numerical framework amongst researchers. It offers a high level programming language to implement new algorithms that support a broad range of high level operations directly on vectors and matrices.

The idea in NumPy is to provide a numerical extension to the Python language that enables the Python language to be both high productive and high performing. NumPy provides not only an API to standardized numerical solvers, but also the option to develop new numerical solvers that are both implemented and efficiently executed in Python, much like the idea behind the Matlab\cite{guide1998mathworks} framework. 

DistNumPy is a new version of NumPy that parallelizes array operations in a manner completely transparent to the user -- from the perspective of the user, the difference between NumPy and DistNumPy is minimal. DistNumPy can use multiple processors through the communication library Message Passing Interface (MPI)\cite{mpi}. However, DistNumPy does not use the traditional single-program multiple-data (SPMD) parallel programming model that requires the user to differentiate between the MPI-processes. Instead, the MPI communication in DistNumPy is fully invisible and the user needs no knowledge of MPI or any parallel programming model. 
The only difference in the API of NumPy and DistNumPy is the array creation routines. DistNumPy allows both distributed and non-distributed arrays to co-exist, the user must specify, as an optional parameter, if the array should be distributed. The following illustrates the only difference between the creation of a standard array and a distributed array:
\lstset{frame=none, xleftmargin=0mm, numbers=none}
\begin{lstlisting}
#Non-Distributed
A = numpy.array([1,2,3])
#Distributed
B = numpy.array([1,2,3], dist=True)
\end{lstlisting}
\lstset{frame=single, xleftmargin=5mm, numbers=left}

\subsection{Views}
\begin{figure}
 \centering
 \includegraphics[width=\linewidth]{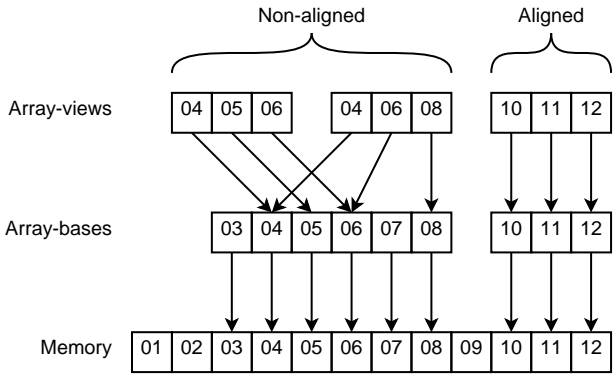}
 \caption{Reference hierarchy between the two array data structures and the main memory. Only the three array-views at top of the hierarchy are visible from the perspective of the user.}
 \label{fig:views}
\end{figure}

NumPy and DistNumPy use identical arrays syntax, which is based on the Python list syntax. The arrays are indexed positionally, 0 through length -- 1, where negative indexes is used for indexing in the reversed order. Like the list syntax in Python, it is possible to index multiple elements. All indexing that represents more than one element returns a view of the elements rather than a new copy of the elements. This means that an array does not necessarily represent a complete, contiguous block of memory. It is possible to have a hierarchy of arrays where only one array represents a complete contiguous block of memory and the other arrays represent a subpart of that memory. DistNumPy implements an array hierarchy where distributed arrays are represented by the following two data structures.
\begin{description}
\item[Array-base] is the base of an array and has direct access to the content of the array in main memory. An array-base is created with all related meta-data when the user allocates a new distributed array, but the user will never access the array directly through the array-base. The array-base always describes the whole array and its meta-data such as array size and data type.
\item[Array-view] is a view of an array-base. The view can re\-present the whole array-base or only a sub-part of the array-base. An array-view can even represent a non-contiguous sub-part of the array-base. An array-view contains its own meta-data that describes which part of the array-base is visible. The array-view is mani\-pulated directly by the user and from the users perspective the array-view is simply a normal contiguous array.
\end{description}
Array-views are not allowed to refer to each other, which means that the hierarchy is flat with only two levels: array-base below array-view. However, multiple array-views are allowed to refer to the same array-base. This two-tier hierarchy is illustrated in Figure \ref{fig:views}. 

\subsection{Data Layout}
\begin{figure}
 \centering
 \includegraphics[width=150px]{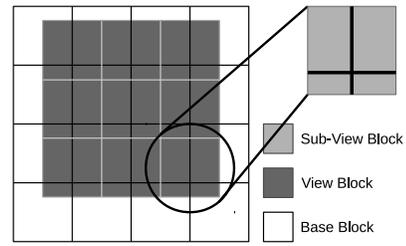}
 \caption{An illustration of the block hierarchy that represents a 2D distributed array. The array is divided into three block-types: Base, View and Sub-View-blocks. The 16 base-blocks make up the base-array, which may be distributed between multiple MPI-processes. The 9 view-blocks make up a view of the base-array and represent the elements that are visible to the user. Each view-block is furthermore divided into four sub-view-blocks, each located on a single MPI-process.}
 \label{fig:view_block}
\end{figure}

The data layout in DistNumPy consists of three kinds of data blocks: base-blocks, view-blocks and sub-view-blocks, which make up a three level abstraction hierarchy (Fig. \ref{fig:view_block}).
\begin{description}
\item[Base-block] is a block of an array-base. It maps directly into one block of memory located on one node. The memory block is not necessarily contiguous but only one MPI-process has exclusive access to the block. Furthermore, DistNumPy makes use of a N-Dimen\-sional Block Cyclic Distribution inspired by High Performance Fortran\cite{Loveman93}, in which base-blocks are distributed across multiple MPI-processes in a round-rob\-in fashion.

\item[View-block] is a block of an array-view, from the perspective of the user a view-block is a contiguous block of array elements. A view-block can span over multiple base-blocks and consequently also over multiple MPI-processes. For a MPI-process to access a whole view-block it will have to fetch data from possibly remote MPI-processes and put the pieces together before accessing the block. To avoid this process, which may cause some internal memory copying, we divide view-blocks into sub-view-block.

\item[Sub-view-block] is a block of data that is a part of a view-block but is located on only one MPI-process. The driving idea is that all array operation is translated into a number of sub-view-block operations.
\end{description}
We will define an \emph{aligned array} as an array that have a direct, contiguous mapping through the block hierarchy. That is, a distributed array in which the base-blocks, view-blocks and sub-view-blocks are identical. A \emph{non-aligned array} is then a distributed array without this property.

\subsection{Universal Function}
An important mechanism in DistNumPy is a concept called a Universal function. A universal function (ufunc) is a function that operates on all elements in an array-view independently. That is, an ufunc is a vectorized wrapper for a function that takes a fixed number of scalar inputs and produces a fixed number of scalar outputs. E.g., addition is an ufunc that takes three array-views as argument: two input arrays and one output array. For each element, the ufunc adds the two input arrays together and writes the result into the output array. Using ufunc can result in a significant performance boost compared to native Python because the computation-loop is implemented in C and is executed in parallel.

Applying an ufunc operation on a whole array-view is semantically equivalent to performing the ufunc operation on each array-view block individually. This property makes it possible to perform a distributed ufunc operation in parallel. 
A distributed ufunc operation consists of four steps:
\begin{enumerate}
\item All MPI-processes determine the distribution of the view-block computation, which is strictly based on the distribution of the output array-view.
\item All MPI-processes exchange array elements in such a manner that each MPI-process can perform its computation locally. 
\item All MPI-processes perform their local computation.
\item All MPI-processes send the altered array elements back to the original locations.
\end{enumerate}

%% file: latency_hiding.tex
%\begin{figure}
% \centering
% \includegraphics[width=200px]{gfx/double_buffering}
% \caption{Flow diagram illustrating double buffering. The $n$'th iteration is expressed with a $n$ and Comm and Comp represents %communication and computation, respectively. $n$++ is an iteration to $n$'s successor.}
% \label{fig:double_buffering}
%\end{figure}

The standard approach to hide communication latency behind computation in message-passing is a technique known as double buffering. The implementation of double buffering is straightforward when operating on a set of data block that all have identical sizes. The communication of one data block is overlapped with the computation of another already communicated data block
% (Fig. \ref{fig:double_buffering}) 
and since the sizes of all the data blocks are identical all iterations are identical.

In DistNumPy, a straightforward double buffering approach works well for ufuncs that operate on aligned arrays, because it translates into communication and computation operations on whole view-blocks, which does not benefit from latency-hiding inside view-blocks. However, for ufuncs that operate on non-aligned arrays this is not the case because the view-block is distributed between multiple MPI-processes. In order to achieve good scalable performance the DistNumPy implementation must therefore introduce latency-hiding inside view-blocks. For example the computation of a view-block in Figure \ref{fig:view_block} can make use of latency-hiding by first initiating the communication of the three non-local sub-view-blocks then compute the local sub-view-block and finally compute the three communicated sub-view-blocks. 

\subsection{The Dependency Graph}
One of the key contributions in this paper is a latency-hiding model that, by maintaining data dependencies between scheduled operations, is able to aggressively initiate communication and lazily evaluate tasks, in order to allow maximal time for the communication to finish before entering a wait state. In this section, we will demonstrate the idea of the model by giving an example of a small 3-point stencil computation implemented in DistNumPy (Fig. \ref{lst:stencil_eq}). For now, we will use a traditional DAG for handling the data dependencies. Later we will describe the implementation of the heuristic that enables us to manage dependencies more efficiently.

Additional, it should be noted that the parallel processes do not need to exchange dependency information since they all have full knowledge of the data distribution.

\begin{figure}
\begin{lstlisting}
import numpy
M = numpy.array([1,2,3,4,5,6],\
                dist=True)
N = numpy.empty((6),dist=True)
A = M[2:] 
B = M[0:4]
C = N[1:5]
C = A + B
\end{lstlisting}
 \caption{This is an example of a small 3-point stencil application.}
 \label{lst:stencil_eq}
\end{figure}

\begin{figure}
 \centering
 \includegraphics[width=150px]{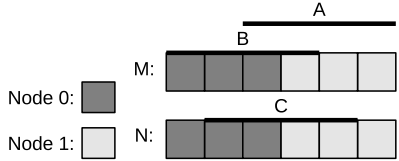}
 \caption{The data layout of the two arrays $M$ and $N$ and the three array-views $A$, $B$ and $C$ in the 3-point stencil application (Fig. \ref{lst:stencil_eq}). The arrays are distributed between two nodes using a block-size of three.}
 \label{fig:stencil_dag1}
\end{figure}

\begin{figure}
 \centering
 \includegraphics[width=\linewidth]{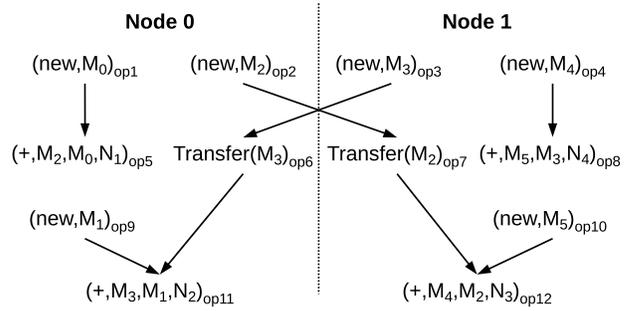}
 \caption{Illustration of a DAG that represents the dependencies in a 3-point stencil application (Fig. \ref {lst:stencil_eq}). The DAG consists of 12 operations, \emph{op1 to op12}, divided between two processes.}
 \label{fig:stencil_dag2}
\end{figure}

Two processes are executing the stencil application and DistNumPy distributes the two arrays, $M$ and $N$, using a block-size of three. This means that three contiguous array elements are located on each process (Fig. \ref{fig:stencil_dag1}). Using a DAG as defined in section \ref{sec:theory}, figure \ref{fig:stencil_dag2} illustrates the dependencies between 12 operations that together constitute the execution. Initially the following six operations are ready:
\begin{equation*}
R := \{op1, op2, op3, op4, op9, op10\}
\end{equation*}
Afterwards, without the need of communication, two more operations $op5$ and $op8$ may be executed. Thus, it is possible to introduce latency-hiding by initiating the communication, $op6$ and $op7$, before evaluating operation $op5$ and $op8$. The amount of communication latency hidden depends on the computation time of $op5$ and $op8$ and the communication time of $op6$ and $op7$. 

We will strictly prioritize between operations based on whe\-ther they involve communication or computation -- giving priority to communication over computation. Furthermore, we will assume that all operations take the same amount of time, which is a reasonable assumption in DistNumPy since it divides array operations into small blocks that often have the same computation or communication time.

\subsection{Lazy Evaluation}
Since Python is an interpreted dynamic programming language, it is not possible to schedule communication and computation operations at compile time. Instead, we introduce lazy evaluation as a technique to determine the communication and computation operations used in the program at runtime.

During the execution of a DistNumPy program all MPI-processes record the requested array operations rather than applying them immediately. The MPI-processes maintain the operations in a convenient data structure and at a later point all MPI-processes apply the operations. The idea is that by having a set of operations to carry out it may be possible to schedule communication and computation operations that have no mutual dependencies in parallel.

We will only introduce lazy evaluation for Python operations that involve distributed arrays. If the Python interpreter encounters operations that do not include DistNumPy arrays, the interpreter will execute them immediately. At some point, the Python interpreter will trigger DistNumPy to execute all previously recorded operation.  This mechanism of executing all recorded operation we will call an \emph{operation flush} and the following three conditions may trigger it.
\begin{itemize}
\item The Python interpreter issues a read from distribut\-ed data. E.g. when the interpreter reaches a branch statement.
\item The number of delayed operations reaches a user-de\-fined threshold.
\item The Python interpreter reaches the end of the program.
\end{itemize}

\subsection{The Dependency System}\label{sec:dep_impl}
The main challenge when introducing lazy evaluation is to implement a dependency system that schedules operations in a performance efficient manner while the implementation keeps the overhead at an acceptable level.

Our first lazy evaluation approach makes use of a DAG-bas\-ed data structure to contain all recorded operations. When an operation is recorded, it is split across the sub-view-blocks that are involved in the operation. For each such operation, a DAG node is created just as in figure \ref{lst:stencil_eq} and \ref{fig:stencil_dag1}.
 
Beside the DAG our dependency system also consist of a \emph{ready queue}, which is a queue of recorded operations that do not have any dependencies. The ready queue makes it possible to find operations that are ready to be executed in the time complexity of $O(1)$.

\paragraph{Operation Insertion}
The recording of an operation triggers an insertion of new node into the DAG. A straightforward approach will simply implement insertion by comparing the new node with all the nodes already located in the DAG. If a dependency is detected the implementation adds an edge between the nodes. The time complexity of such an implementation is $O(n)$ where $n$ is the number of operation in the DAG and the construction of the complete DAG is $O(n^2)$.

%In worse case this approach is optimal because the longest path through the DAG is always $n$. However, we have introduced some heuristics to speed up the common case. 

\paragraph{Operation Flush}
To achieve good performance the operation flush implementation must maximize the amount of communication that it is overlapped by computation. Therefore, the flush implementation initiate communication at the earliest point in time and only do computation when all communication has been initiated. Furthermore, to make sure that there is progress in the MPI layer it checks for finished communication in between multiple computation operations. The following is the flow of our operation flush algorithm:
\begin{enumerate}
\item Initiate all communication operations in the ready qu\-eue.
\item Check in a non-blocking manner if some communication operations have finished and remove finished communication operations from the ready queue and the DAG. Furthermore, register operations that now have no dependencies into the ready queue.
\item If there is only computation operations in the ready queue execute one of them and remove it from the ready queue and the DAG.
\item Go back to step one if there are any operations left in the ready queue else we are finished.
\end{enumerate}
The algorithm maintains the following three invariants:
\begin{enumerate}
\item All operations, that are ready, are located in the ready queue.
\item We only start the execution of a computation node when there is no communication node in the ready queue.
\item We only wait for communication when the ready queue has no computation nodes.
\end{enumerate}

\subsubsection{Deadlocks}

\begin{figure}
 \centering
 \includegraphics[width=\linewidth]{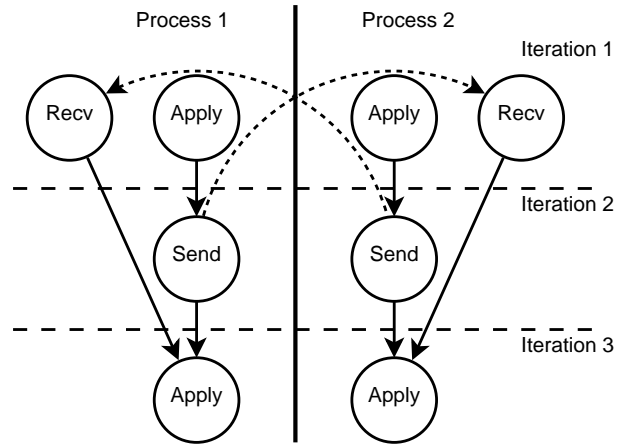}
 \caption{Illustration of the na\"\i ve evaluation approach. The result is a deadlock in the first iteration since both processes are waiting for the receive-node to finish, but that will never happen because the matching send-node is in second iteration.}
 \label{fig:DependencyGraph_Deadlock}
\end{figure}

%\begin{figure}
% \centering
% \includegraphics[width=\linewidth]{gfx/DependencyGraph_DeadlockFree}
% \caption{Illustration of a deadlock-free evaluation of the dependency graph in figure \ref{fig:DependencyGraph_Deadlock}. Each MPI-process evaluates as much as possible before waiting for any communication.}
% \label{fig:DependencyGraph_DeadlockFree}
%\end{figure}

To avoid deadlocks a MPI-process will only enter a blocking state when it has initiated all communication and finished all ready computation. This guaranties a deadlock-free execution but it also reduces the flexibility of the execution order. Still, it is possible to check for finished communication using non-blocking functions such as \texttt{MPI\_Testsome()}.

The na\"\i ve approach to evaluate a DAG is simply to first evaluate all nodes that have no dependencies and then remove the evaluated nodes from the graph and start over -- similar to the traditional BSP model. However, this approach may result in a deadlock as illustrated in figure \ref{fig:DependencyGraph_Deadlock}. 
%Figure \ref{fig:DependencyGraph_DeadlockFree} illustrate the same DAG executed using our approach where a MPI-process evaluates as much as possible before entering a blocking state.

\subsubsection{Dependency Heuristic}
Experiments with lazy evaluation using the DAG-based data structure shows that the overhead associated with the creation of the DAG is very time consuming and becomes the dominating performance factor. We therefore introduce a heuristic to speed up the common case. We base the heuristic on the following two observations:
\begin{itemize}
\item In the common case, a scientific DistNumPy application spreads a computation evenly between all sub-view-blocks in the involved arrays.
\item Operation dependencies are only possible between sub-view-blocks that are part of the same base-block.
\end{itemize}
The heuristic is that instead of having a DAG, we introduce a prioritized operation list for each base-block. The assumption is that, in the common case, the number of operations associated with a base-block is manageable by a linked list.

%\subsubsection{Dependency-lists}
We implement the heuristic using the following algorithm. A number of operation-nodes and access-nodes represent the operation dependencies. The operation-node contains all information needed to execute the operation on a set of sub-view-blocks and there is a pointer to an access-node for each sub-view-block. The access-node represents memory access to a sub-view-block, which can be either reading or writing. E.g., the representation of an addition operation on three sub-view-blocks is two read access-nodes and one write access-node (Fig. \ref{fig:dependency_system_term}).

Our algorithm places all access-nodes in dependency-lists based on the base-block that they are accessing. When an operation-node is recorded each associated access-node is inserted into the dependency list of the sub-view-blocks they access. Additionally, the number of accumulated dependencies the access-nodes encounter is saved as the operation-node's reference counter.

All operation-nodes that are ready for execution have a reference count of zero and are in the ready queue. Still, they may have references to access-nodes in dependency-lists -- only when we execute an operation-node will we remove the associated access-nodes from the dependency-lists. Following the removal of an access-node we traverse the dependency-list and for each depending access-node we reduce the associating reference counter by one. Because of this, the reference counter of another operation-node may be reduced to zero, in which case we move the operation-node to the ready queue and the algorithm starts all over.

\begin{figure}
 \centering
 \includegraphics[width=170px]{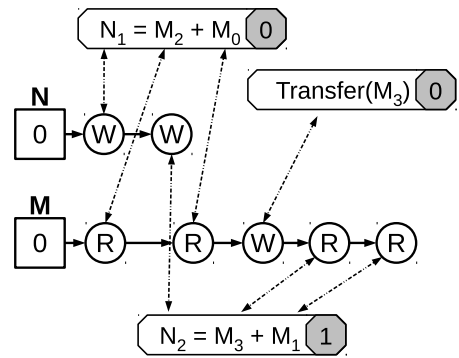}
 \caption{Illustration of the dependency system when executing the 3-point stencil in figure \ref{lst:stencil_eq}, \ref{fig:stencil_dag1} and \ref{fig:stencil_dag2}. The illustration is a snapshot of the dependency system on node 0 after the creation of all the arrays. Note that since the block size is three, node 0 only has one block of each array.}
 \label{fig:dependency_system}
\end{figure}

Figure \ref{fig:dependency_system_term} goes through all the structures that make up the dependency system and figure \ref{fig:dependency_system} illustrates a snapshot of the dependency system when executing the 3-point stencil application. 

\def\imagetop#1{\vtop{\null\hbox{#1}}}
\begin{figure}
 \centering
\begin{tabular}{cp{165px}}
\imagetop{\includegraphics[width=30px]{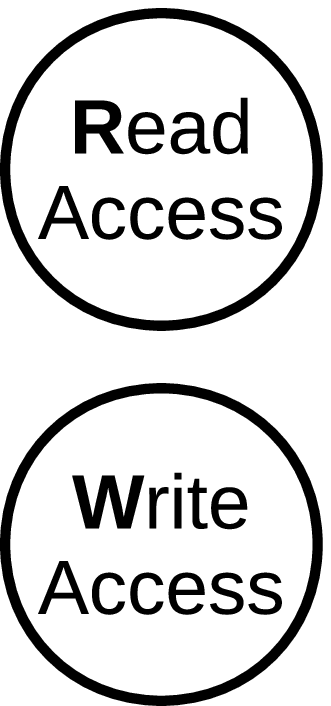}} & \vspace{1px} All access-nodes that access the same base-block are linked together in a de\-pen\-dency-list. The order of the list is simply based on the time of node insertion (descending order). Additionally an access-node contains the information needed to determine whether it depends on other access-nodes.\newline \\
\imagetop{\includegraphics[width=50px]{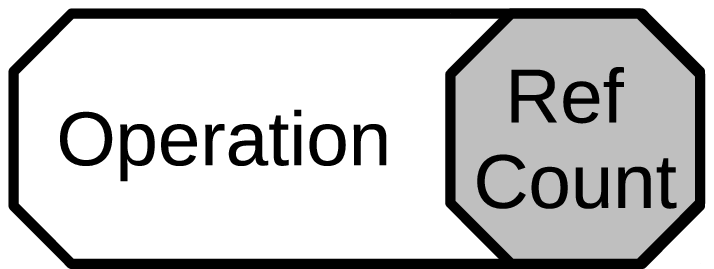}} & An operation-node has a pointer to all access-nodes that are involved in the operation. Attached to an operation is a re\-ference counter that specifies the number of dependencies associated with the op\-eration. When the counter reaches zero the operation is ready for execution. At some point when the operation has been executed the operation-node and all access-nodes are completely remove from the dependency system.\\
\imagetop{\includegraphics[width=28px]{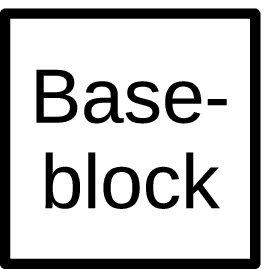}} & \vspace{3px} A base-block simply contains a pointer to the first access-node in the dependency-list.\\
\end{tabular}
 \caption{The structures used in the dependency system.}
 \label{fig:dependency_system_term}
\end{figure}

%% file: benchmark.tex
\begin{table}
\caption{Hardware specifications}
\centering
\begin{tabular}{|ll|}
\hline
CPU & Intel Xeon E5345 \\
CPU Frequency & 2.33 GHz \\
CPU per node & 2 \\
Cores per CPU & 4 \\
Memory per node & 16 GB \\
Number of nodes & 16 \\
Network & Gigabit Ethernet\\
\hline
\end{tabular}
\label{tab:specs}
\end{table}
%https://fyrkat.grid.aau.dk/wiki/index.php/User_pages

To evaluate the performance impact of the latency-hiding model introduced in this paper, we will conduct performance benchmark using DistNumPy and NumPy\footnote{NumPy version 1.3.0}.  The benchmark is executed on an Intel Core 2 Quad cluster (Table \ref{tab:specs}) and for each application we calculate the speedup of DistNumPy compared to NumPy. The problem size is constant though all the executions, i.e. we are measuring strong scaling.  To measure the performance impact of the latency-hiding, we use two different setups: one with latency-hiding and one that uses blocking communication. For both setups we measured the time spent on waiting for communication, i.e. the communication latency not hidden behind computation.

In this benchmark we utilizes the cluster in a \emph{by node} fashion. That is, from one to sixteen CPU-cores we start on MPI-process per node (no SMP) and above sixteen CPU-cores we start multiple MPI-processes per node. The MPI library used throughout this benchmark is OpenMPI\footnote{OpenMPI version 1.5.1}.

\begin{figure}
 \centering
 \includegraphics[angle=270, width=\linewidth]{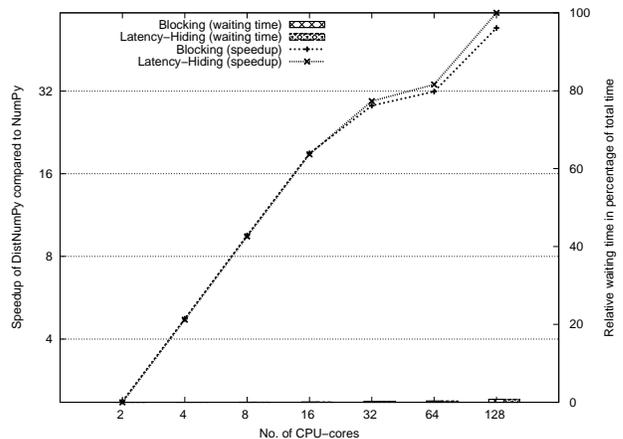}
 \caption{Speedup of the Fractal application.}
 \label{fig:fractal}
\end{figure}

\begin{figure}
 \centering
 \includegraphics[angle=270, width=\linewidth]{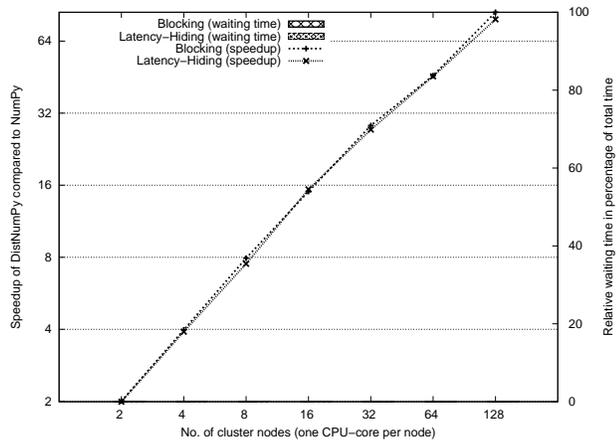}
 \caption{Speedup of the Black-Scholes application.}
 \label{fig:BlackScholes}
\end{figure}

\begin{figure}
 \centering
 \includegraphics[angle=270, width=\linewidth]{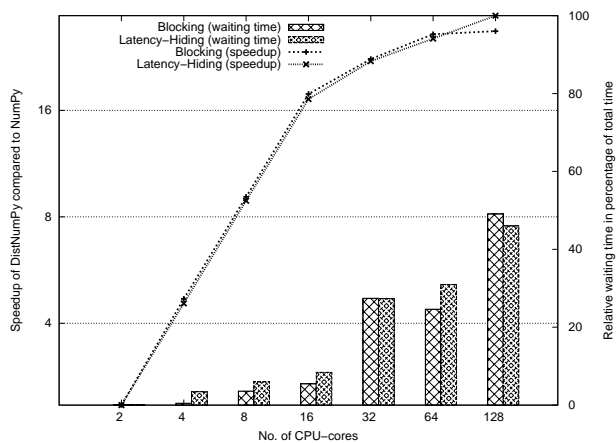}
 \caption{Speedup of the N-body application.}
 \label{fig:nbody}
\end{figure}

\begin{figure}
 \centering
 \includegraphics[angle=270, width=\linewidth]{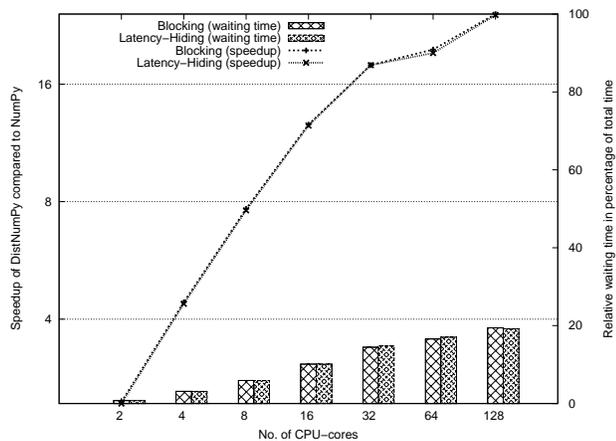}
 \caption{Speedup of the kNN application.}
 \label{fig:knn}
\end{figure}

\begin{figure}
 \centering
 \includegraphics[angle=270, width=\linewidth]{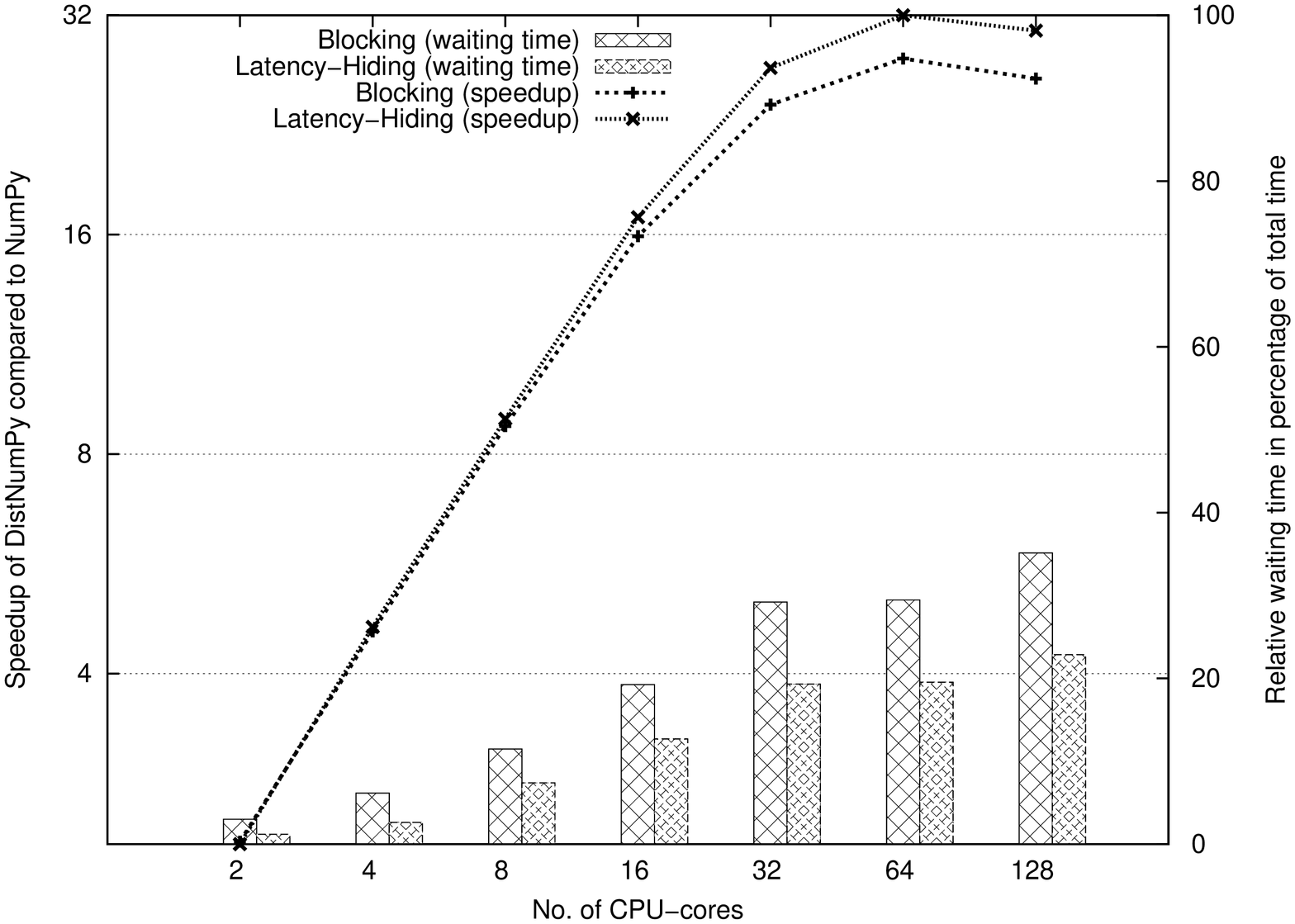}
 \caption{Speedup of the Lattice Boltzmann 2D application.}
 \label{fig:lbm2d}
\end{figure}

\begin{figure}
 \centering
 \includegraphics[angle=270, width=\linewidth]{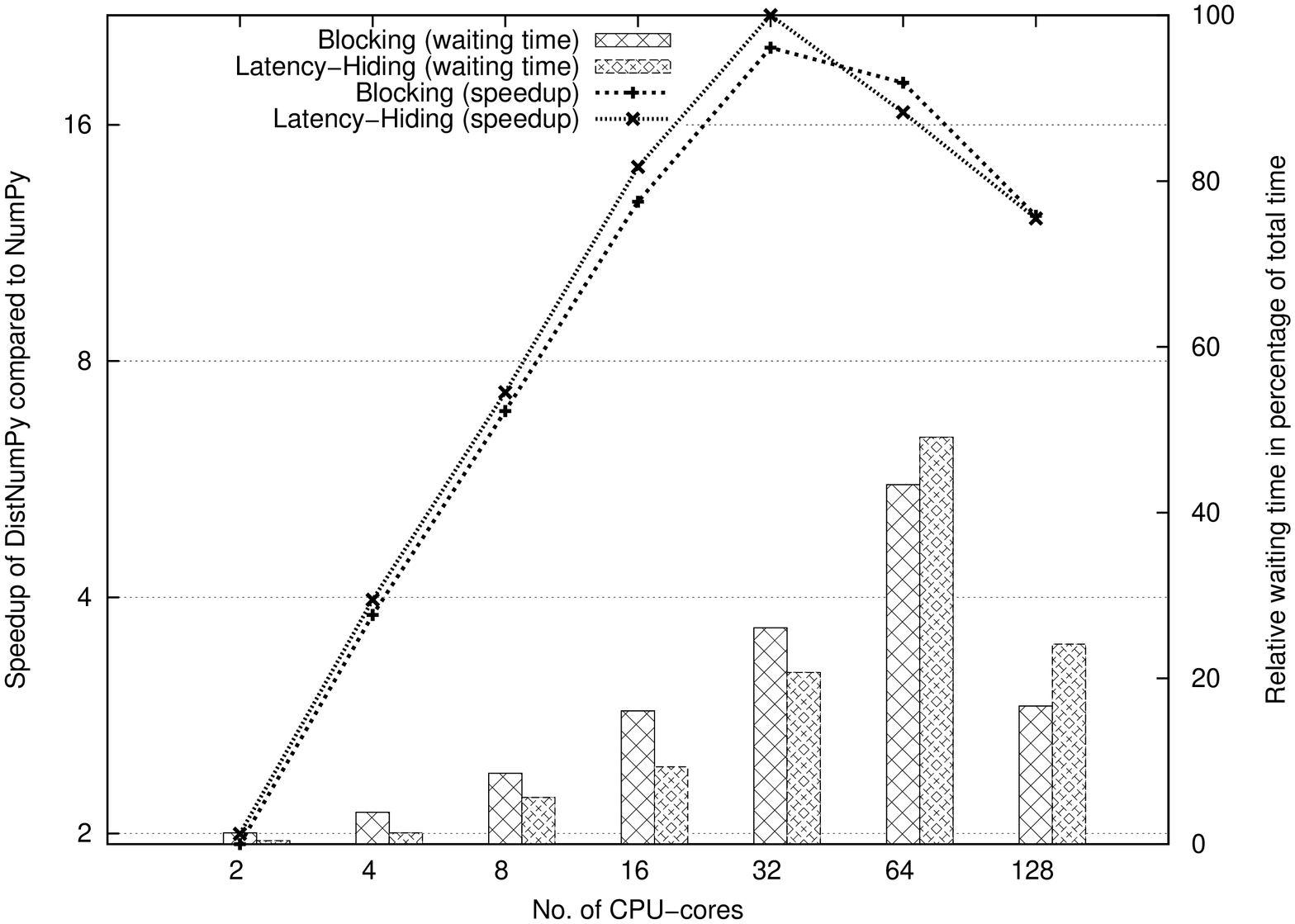}
 \caption{Speedup of the Lattice Boltzmann 3D application.}
 \label{fig:lbm3d}
\end{figure}

\begin{figure}
 \centering
 \includegraphics[angle=270, width=\linewidth]{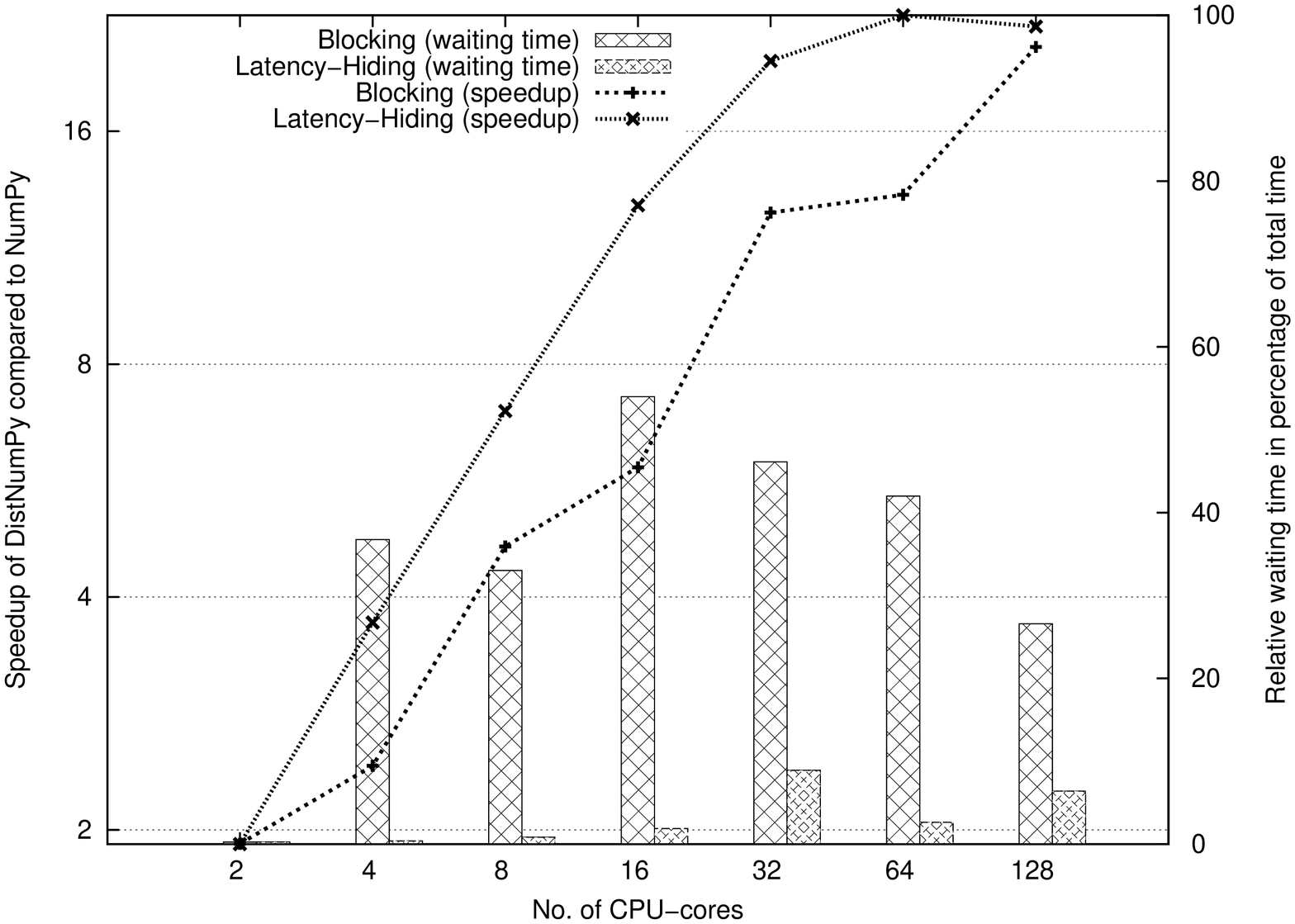}
 \caption{Speedup of the Jacobi application.}
 \label{fig:jacobi}
\end{figure}

\begin{figure}
 \centering
 \includegraphics[angle=270, width=\linewidth]{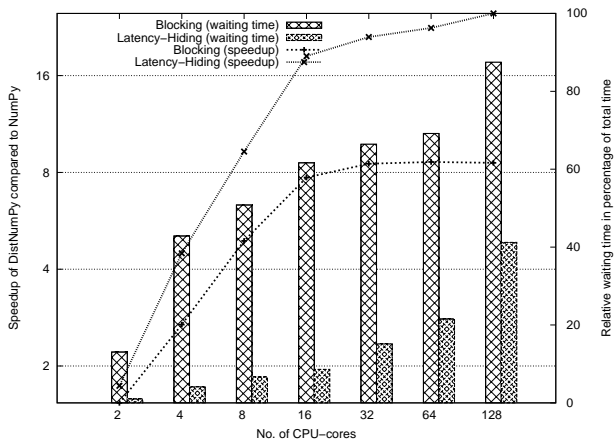}
 \caption{Speedup of the Jacobi Stencil application.}
 \label{fig:jacobi_stencil}
\end{figure}

The benchmark consists of the following eight Python applications.
\begin{description}
\item[Fractal] Computation of the Mandelbrot Set. From a Num\-Py tutorial written by Walt\cite{Walt08_numpy_tutorial} (Fig. \ref{fig:fractal}).
\item[Black-Scholes] Computation of the Black-Scholes model\cite{black1973} implemented in NumPy (Fig. \ref{lst:BlackScholes} and \ref{fig:BlackScholes}).
%\item[Monte Carlo] Approximating Pi using Monte Carlo simulation. The implementation is a translation of the Monte Carlo simulation included in the benchmark suite SciMark 2.0\cite{SciMark}, which is written in Java (Fig. \ref{fig:MonteCarlo}). 

Both \textbf{Fractal} and \textbf{Black-Scholes} are embarrassingly parallel applications and we expect that latency-hiding will not improve the performance.

\item[N-body] A Newtonian N-body simulation that uses a na\"\i ve algorithm that computes all body-body interactions. The NumPy implementation is a translation of a MATLAB application by Casanova\cite{assignmentNbody} (Fig. \ref{fig:nbody}).
\item[kNN] A na\"\i ve implementation of a k nearest neighbor search (Fig. \ref{fig:knn}). 

The \textbf{N-body} and \textbf{kNN} applications have a computation complexity of $O(n^2)$. This indicates that the two applications should have good scalability even without latency-hiding.

\item[Lattice Boltzmann 2D] Lattice Boltzmann model of channel flow in 2D using the D2Q9 model. It is a translation of a MATLAB application by Latt\cite{Latt06_lbm2d} (Fig. \ref{fig:lbm2d}).
\item[Lattice Boltzmann 3D] Lattice Boltzmann model of a flu\-id in 3D using the D3Q19 model. It is a translation of a MATLAB application by Haslam\cite{Haslam06_lbm3d} (Fig. \ref{fig:lbm3d}). 

The two \textbf{Lattice Boltzmann} applications have a computation complexity of $O(n)$. More communication is therefore needed and we expect that latency-hid\-ing will improve the performance.

\item[Jacobi] The Jacobi method is an algorithm for determining the solutions of a system of linear equations. It is an iterative method that uses a spitting scheme to approximate the result (Fig. \ref{fig:jacobi}). 
\item[Jacobi Stencil] In this benchmark, we have implemented the Jacobi method using stencil operations rather than matrix row operations (Fig. \ref{lst:jacobi_stencil} and \ref{fig:jacobi_stencil}). 

The two \textbf{Jacobi} applications also have a computation complexity of $O(n)$. However, the constant associated with $n$ is very small, e.g. to compute one element in the Jacob Stencil application four adjacent elements are required. We expect latency-hiding to be very important for good scalability.
\end{description}

\lstset{numbers=none}
\begin{figure}
\begin{lstlisting}
# Black Scholes Function
#  S: Stock price
#  X: Strike price
#  T: Years to maturity
#  r: Risk-free rate
#  v: Volatility
def BlackScholes(CallPutFlag,S,X,T,r,v):
 d1 = (log(S/X)+(r+v*v/2.)*T)/(v*sqrt(T))
 d2 = d1-v*sqrt(T)
 if CallPutFlag=='c':
   return S*CND(d1)-X*exp(-r*T)*CND(d2)
 else:
   return X*exp(-r*T)*CND(-d2)-S*CND(-d1)
\end{lstlisting}
 \caption{This is the Black Sholes Function in the \textbf{Black-Scholes} benchmark where \texttt{CND} is the cumulative normal distribution. Note that there is no source code difference between a parallel and a sequential version -- it is regular Python/Numpy source code.}
 \label{lst:BlackScholes}
\end{figure}

\begin{figure}
\begin{lstlisting}
cells = full[1:-1, 1:-1]
up    = full[0:-2, 1:-1]
down  = full[2:  , 1:-1]
left  = full[1:-1, 0:-2]
right = full[1:-1, 2:  ]
while epsilon < delta:
  work[:] = cells
  work += 0.2 * (up+down+left+right)
  diff = absolute(cells - work)
  delta = sum(diff)
  cells[:] = work
\end{lstlisting}
 \caption{This is the kernel in the \textbf{Jacobi Stencil} benchmark. First, we define a view of the full array for each point in the stencil, five in this case, and then we apply the stencil until it converges. Note that there is no source code difference between a parallel and a sequential version -- it is regular Python/Numpy source code.}
 \label{lst:jacobi_stencil}
\end{figure}

\subsection{Discussion}
Overall, the benchmarks show that DistNumPy has both good performance and scalability. However, the scalability is somewhat worsening at 32 CPU-cores and above, which correlates with the use of multiple CPU-cores per node. Because of this distinct performance profile, we separate the following discussion into results executed on one to sixteen CPU-cores (one CPU-core per node) and the results executed on 32 CPU-cores to 128 CPU-cores (multiple CPU-cores per node).

\subsubsection{One to Sixteen CPU-cores}
The benchmarks clearly shows that DistNumPy has both good performance and scalability. Actually, half of the Py\-thon applications achieve super-linear speedup at sixteen CPU-cores. This is possible because DistNumPy, opposed to NumPy, will try to reuse memory allocations by lazily de-allocating arrays. DistNumPy uses a very na\"\i ve algorithm that simply checks if a new array allocation is identical to a just de-allocated array. If that is the case one memory allocation and de-allocation is avoided.

In the two embarrassingly parallel applications, \textbf{Fractal} and \textbf{Black-Scholes}, we see very good speedup as expected. Since the use of communication in both applications is almost non-existing the latency-hiding makes no difference. The speedup achieved at sixteen CPU-cores is 18.8 and 15.4, respectively.

The two na\"\i ve implementations of \textbf{N-body} and \textbf{kNN} do not benefit from latency-hiding. In \textbf{N-body} the dominating operations are matrix-multiplications, which is a native operation in NymPy and in DistNumPy implemented as specialized operations using the parallel algorithm SUMMA\cite{SUMMA_GeijnW97} and not as a combination of ufunc operations. Since both the latency-hiding and the blocking execution use the same SUMMA algorithm the performance is very similar. However, because of the overhead associated with latency-hiding, the blocking execution performs a bit better. The speedup achieved at sixteen CPU-cores is 17.2 with latency-hiding and 17.8 with blocking execution. Similarly, the performance difference between latency-hiding and blocking in \textbf{kNN} is minimal -- the speedup achieved at sixteen CPU-cores is 12.5 and 12.6, respectively. The relatively modest speedup in \textbf{kNN} is the result of poor load balancing. At eight and sixteen CPU-cores the chosen problem is not divided evenly between the processes.

Latency-hiding is somewhat beneficial to the two \textbf{Lattice Boltzmann} applications. The waiting time percentage on sixteen CPU-cores goes from 19\% to 13\% in \textbf{Lattice Boltzmann 2D}, and from 16\% to 9\% in \textbf{Lattice Boltzmann 3D}. However, the overall impact on the speedup is not that great, primarily because of the overhead associated with latency-hiding.

Finally, latency-hiding introduces a drastically improved performance to the two communication intensive applications \textbf{Jacobi} and \textbf{Jacobi Stencil}. The waiting time percentage going from 54\% to 2\% and from 62\% to 9\%, respectively. Latency-hiding also has a major impact on the overall speedup, which goes from 5.9 to 12.8 and from 7.7 to 18.4, respectively. In other words latency-hiding more than doubles the overall speed\-up and CPU utilization.

\subsubsection{Scaling above sixteen CPU-cores}
\begin{figure}
 \centering
 \includegraphics[angle=270, width=0.95\linewidth]{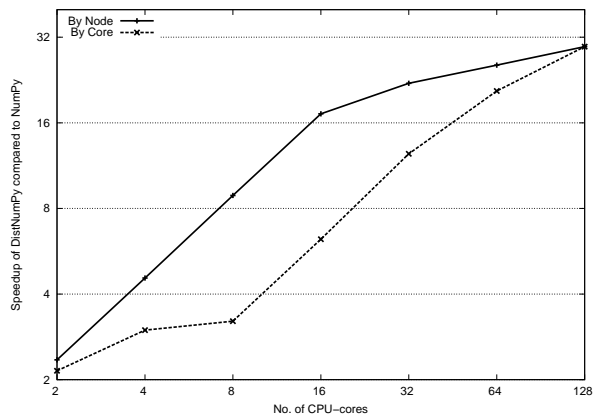}
 \caption{Speedup of the N-body application that compares \emph{by node}, in which the maximum number of CPU-cores is used, and \emph{by core}, in which the minimum number of nodes is used. Note that the \emph{by node} graph is identical to the \emph{latency-hiding} graph in figure \ref{fig:nbody}.}
 \label{fig:nbody_smp}
\end{figure}

Overall, the performance is worsening at 32 CPU-cores -- particular at 64 CPU-cores and above where the CPU utilization is below 50\%. One reason for this performance degradation is the characteristic of strong scaling. In order to have considerable more data blocks than MPI-processes, the size of the data distribution blocks decreases as the number of executing CPU-cores increases. Smaller data blocks reduces the performance since the overhead in DistNumPy is proportional with the size of a data block. 

However, smaller data blocks are not enough to justify the observed performance degradation. The von Neumann bottleneck\cite{Backus78} associated with main memory also hinder scalability. This is evident when looking at Figure \ref{fig:nbody_smp}, which is a speedup graph of \textbf{N-body} that compares \emph{by node} and \emph{by core} scaling. At eight CPU-cores, both result uses identical data distribution and block size, but the performance when only using one CPU-core per node is clearly superior to using all eight CPU-cores on one node. 

A NumPy application will often use ufuncs heavily, which makes the application vulnerable to the von Neumann bottleneck. The problem is that multiple ufunc operations are not pipelined in order to utilize cache locality. Instead, NumPy will compute a single ufunc operation at a time. This problem is also evident in DistNumPy since our latency-hiding model only address communication latency and not memory latency.

%% file: futurework.tex
The latency-hiding model introduced in this paper focuses on communication latency. However, the result from out benchmarks shows that memory latency is another aspect that is important for good scalability -- particular when utilizing shared memory nodes. 

One approach to address this issue is to extend our latency-hiding model with cache locality optimization. The scheduler will have to prioritize computation operations that are likely to be in the cache. One heuristic to accomplish this is to sort the operations in the ready queue after the last time the associated data block has been accessed.

Another approach is to merge calls to ufuncs, that operate on common arrays, together in one joint operation and thereby make the joint operation more CPU-intensive. If it is possible to merge enough ufuncs together the application may become CPU bound rather than memory bound.

%Vi kunne ogsaa bruge JIT ala weave.blitz

Introducing Hybrid Programming could also be a solution to the problem. In order to utilize hybrid architectures, \cite{kristensen2011_gpaw} shows that shared and distributed memory programming can improve the overall performance and scalability.

%% file: conclusions.tex
While automatic parallelization for distributed memory architectures cannot hope to compete with a manually parallelized version, the productivity that comes with automatic parallelization still makes the technique of interest to a user who only runs a code a few times between changes. For applications that are embarrassingly parallel or applications where the computational complexity is $O(n^2)$ or higher, it is fairly straight forward to manage the communication for automatic parallelization. However, for common kernels the complexity is $O(n \log(n))$ or even $O(n)$ and here the application of latency-hiding techniques is essential for performance.

In this work we have presented a scheme for managing la\-tency-hiding, that is based on the assumption that splitting the work in more blocks than there are processors will allow us to aggressively communicate data-blocks between nodes, while at the same time processing operations that require no external data-blocks. The same dependency analysis may be done without a division into data-blocks, but the blocking approach allows us to maintenance of a full DAG, an operation that is known to be costly, and replace the DAG with a number of ordered linked lists, to which access is done in linear time.

We implement the model in Distributed Numerical Python, DistNumPy, a programming framework that allows linear algebra operations expressed in NumPy to be executed on distributed memory platforms and this is without any effort towards parallelization from the programmer. 

A selection of eight benchmarks show that the system, as predicted, does not improve the performance of embarrassingly parallel applications or applications with complexity $O(n^2)$ or higher. For applications with lower complexity the benefit from automatic latency-hiding is highly dependent on the relationship between the amount of data that needs to be transferred and the cost of updating the individual elements. In the Lattice Boltzmann codes, both 2D and 3D, the version without latency-hiding does quite well, simply because the operation of updating a data-point is time consuming enough to amortize the communication latency. However, when the cost of communication becomes more significant, as in a Jacobi solver and stencil-based Jacobi solver, the automatic latency-hiding significantly improve the performance. The performance of the stencil-based Jacobi-solver improves from a speedup of 7.7 to 18.4 at sixteen processors and 8.6 to 25.0 at 128 processors, compared to standard sequential NumPy. This is matched by the fact that the time spend on waiting for communication drops from 62\% to 9\% and 87\% to 41\%, respectively, with the introduction of latency-hiding. 

Overall, the conclusion is that managing latency-hiding at runtime is fully feasible and makes automatic parallelization feasible for a number of applications where manual parallelization would otherwise be required. The most obvious target is the large base of stencil-based algorithms.